\documentclass[aps,prl,twocolumn,groupedaddress,showpacs]{revtex4-1}
 \usepackage{graphicx}
 \usepackage{dcolumn}
 \usepackage{bm}
 \usepackage{amsmath}
 \usepackage{color}
\begin{document}


\title{Evidence of the side jump mechanism in the anomalous Hall effect in paramagnets}


\author{Yufan Li}
\author{Gang Su}
\author{Dazhi Hou}
\author{Li Ye}
\author{Yuan Tian}
\author{Jianli Xu}
\author{ Xiaofeng Jin}
\email[Corresponding author,email: ]{xfjin@fudan.edu.cn}
\affiliation{ State Key Laboratory of Surface Physics and Department of Physics, Fudan University, Shanghai
200433, China }


\date{\today}

\begin{abstract}

Persistent confusion has existed between the intrinsic (Berry curvature) and the side jump mechanisms of anomalous Hall effect (AHE) in ferromagnets. We provide unambiguous identification of the side jump mechanism, in addition to the skew scattering contribution in epitaxial paramagnetic Ni$_{34}$Cu$_{66}$ thin films, in which the intrinsic contribution is by definition excluded. Furthermore, the temperature dependence of the AHE further reveals that the side jump mechanism is dominated by the elastic scattering.
\end{abstract}


\maketitle

The anomalous Hall effect (AHE) has been an intriguing spin-dependent transport phenomenon in condensed matter physics with its microscopic origin remaining unsettled \cite{NagaosaRev2010}.
Experimental observations generally found a power law relationship between the anomalous Hall resistivity and the longitudinal resistivity of $\rho_{AH}\sim\rho_{xx}^{\xi}$, where $\xi$ can be 1 or 2 according to the different and competing proposed mechanisms \cite{KarplusLuttingerIntrinsic,SmitSkew,BergerSideJump}. Ascribing certain mechanisms to the experimental observations is challenging and often results in controversy.

Karplus and Luttinger (KL) proposed the first microscopic theory later known as the intrinsic mechanism \cite{KarplusLuttingerIntrinsic}. It indicates a scaling exponent of $\xi=$ 2, which could well explain some experimental results such as the AHE of iron \cite{Kooi}, but fails to account for the AHE in other metals such as nickel \cite{SmitSkew1st,Ni_bulk}, and dilute alloys \cite{AFertleft-right,FertCeLa,FertCe,FertAgAuAl,FertHallrev}.
On the other hand, Smit argued that the inevitable impurity scattering should dominate the AHE and he proposed an extrinsic mechanism called the skew scattering which gives $\xi=$ 1 \cite{SmitSkew1st,SmitSkew}, in agreement with numerous experiments \cite{AFertleft-right,FertCeLa,FertCe,FertAgAuAl,FertHallrev}, but not others such as iron. Fifteen years later Berger proposed another extrinsic mechanism known as the side jump with $\xi=$ 2 \cite{BergerSideJump}. Since then it appears that the extrinsic mechanisms alone can adequately explain the AHE, and for decades the idea of the KL theory was put aside \cite{spinHall,Coey_book,Bruno2001}. However, the intrinsic mechanism was revived after the KL theory was reformulated by the modern Berry phase language  \cite{QNiuBerryPhase}, which can quantitatively predict the intrinsic anomalous Hall conductivity for a given material by the first-principles calculation.
Reasonable agreement was found between the calculated intrinsic AHE and the observed overall AHE in a wide range of materials including ferromagnetic semiconductors \cite{NiuMacBerryPhase}, complex oxides \cite{FangMonopole}, transition metals \cite{bccFeCal} and their alloys \cite{zengMn5Si3}. These developments reestablish the intrinsic mechanism. In short, AHE has been controversial for over fifty years with opinions swaying back and forth, yet without consensus.

The current trend seems to favor again the intrinsic mechanism \cite{NagaosaRev2010,DiXiaoRev}.
Most of the experimental observations of $\xi=$ 2 which was once ascribed to the side jump but now attributed to the intrinsic mechanism. This leaves little room for the side jump mechanism \cite{DiXiaoRev}, and even creates doubts whether it can be really detected in experiments \cite{DoesSJexist}. Although more recently an extra $\xi=$ 2 term in addition to the intrinsic mechanism was identified experimentally in Fe implying the presence of the side jump \cite{FeAHE}. There are also attempts to compare the measured $\xi=$ 2 term with first principles calculations and then to obtain the side jump contribution as the difference \cite{sidejumpexcal,smzhou,SJabinitio}. To date, there is no direct experimental verification of the very existence of the side jump.

It is well known that the Berry curvature of the Bloch electrons or the KL intrinsic mechanism could contribute to the AHE only in a ferromagnetic system, where the spin-up and spin-down electrons are unequally populated at the Fermi level.
In principle an external magnetic field might induce a nonzero Berry curvature even in paramagnetic systems. However this contribution to the Hall effect is linearly proportional to the magnetic field, thus experimentally indistinguishable with the ordinary Hall effect. Therefore when the AHE is obtained after subtracting the ordinary Hall contribution, such a possible intrinsic component to the AHE is ineluctably excluded. On the other hand, the left-right asymmetry in the scattering of electrons induced by the magnetic moments can exist not only in ferromagnets but in paramagnets as well, giving rise to the extrinsic AHE, as clearly shown by Fert \emph{et al.}, \cite{AFertleft-right}. They studied intensively the AHE in various paramagnetic systems with very diluted impurities, where only the skew scattering (or the linear resistivity term) was found to be relevant \cite{FertCeLa,FertCe,FertAgAuAl,FertPt,FertHallrev}.  It is evident today that in such a clean limit with very low resistivity, even if the side jump or the quadratic resistivity term exists it would be negligibly small comparing to the skew scattering term \cite{Nagaosa3region,Tokura3region,NagaosaRev2010}.
Accordingly, in order to identify the side jump mechanism it is more desirable to explore in paramagnetic but moderately "dirty" systems with higher residual resistivity, in which the quadratic resistivity term is comparable to the skew scattering term. When a quadratic resistivity term can be unambiguously realized in such systems, there are no other alternatives but the side jump.

In this work, we investigated the AHE in Ni$_{0.34}$Cu$_{0.66}$ alloy, which is paramagnetic with magnetic moments. We used epitaxial Ni$_{0.34}$Cu$_{0.66}$ thin films of the same composition, but the resistivity can be tuned by varying the film thickness and temperature. The result clearly show a quadratic resistivity term besides the linear skew scattering term,
 and therefore identifies directly and unambiguously the very existence of the side jump mechanism.
Furthermore, the temperature dependence of the AHE shows that the side jump contribution $\rho_{AH}^{sj}$ scales with the residual resistivity $\rho_{xx0}$ rather than the total resistivity $\rho_{xx}$, which presumably implies that the phonon contribution to the side jump should be negligible compared to that by the static defects in the material.

\begin{figure}
\includegraphics[width=8.5cm]{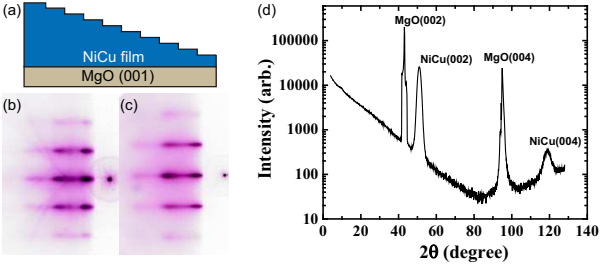}
\caption{(color online) (a) An illustration of the thickness steps of the paramagnetic Ni$_{1-x}$Cu$_{x}$ film. (b) The RHEED pattern of the MgO(001) substrate. (c) The RHEED pattern of the Ni$_{34}$Cu$_{66}$ /MgO(001) film. (d) Grazing-incident X-ray diffraction of the 10 nm-thick film of Ni$_{34}$Cu$_{66}$ /MgO(001).}
\end{figure}

Ni$_{1-x}$Cu$_{x}$ film of thickness varying from 5~nm to 15~nm was grown on MgO(001) substrate at 150~K by molecular beam epitaxy. A 5~nm-thick MgO protecting layer was subsequently deposited onto the sample before taken out from the ultrahigh vacuum to avoid the oxidation during the transport measurement. More details of the experimental setting can be found elsewhere \cite{LFYin,FeAHE,XiaoBi}. It is well established that the Curie temperature $T_c$ of ferromagnetic Ni$_{1-x}$Cu$_{x}$ alloys drop to 0 K at the composition of $x=$42\% \cite{NiCuXc}, therefore we choose the composition of $x=$34\% to ensure its paramagnetism.
A series of face-centered-cubic (fcc) Ni$_{0.34}$Cu$_{0.66}$ films with various well-controlled thicknesses are deposited on a single substrate as illustrated in Fig. 1(a), by employing a shadow mask technique \cite{NiAHE}. The single crystalline nature is clearly seen from the reflection high-energy electron diffraction (RHEED) pattern of [Fig.~1(c)] and the high crystalline quality is further verified by the grazing-incident X-ray diffraction (XRD) [Fig.~1(d)]. Because of the distinct lattice mismatch between fcc-Ni and fcc-Cu as well as the total miscibility of Ni and Cu in Ni$_{1-x}$Cu$_{x}$ for the whole $x$ range between 0 and 1, noticeable phase segregated Ni clusters can be excluded by such high quality single crystallinity shown by RHEED and XRD. As expected, no spontaneous magnetization or any trace of super-paramagnetism can be detected by the SQUID magnetometery (not shown). The sample was further patterned into standard Hall bars by photolithography for the transport measurement.

Fig. 2(a) shows the temperature dependence of the longitudinal resistivity ($\rho_{xx}$) of the Ni$_{34}$Cu$_{66}$ thin films for various thicknesses.  The purpose of tuning $\rho_{xx}$ via the thickness and the temperature is two-fold. (i) the tuning via the film thickness at fixed temperature $T=5$~K can separate the skew scattering from the side jump by distinguishing the linear and quadratic resistivity terms \cite{HouCobalt}; (ii) the tuning via the temperature for a film with certain thickness will help clarify whether the scaling of the side jump is $\rho_{AH}^{sj}\propto\rho_{xx0}^2$ or $\rho_{AH}^{sj}\propto\rho_{xx}^2$, which in return tells whether the phonon-induced inelastic scattering contributes significantly to the side jump (the latter) or not (the former). These two issues will be tackled in the following.

\begin{figure}
\includegraphics[width=6.5cm]{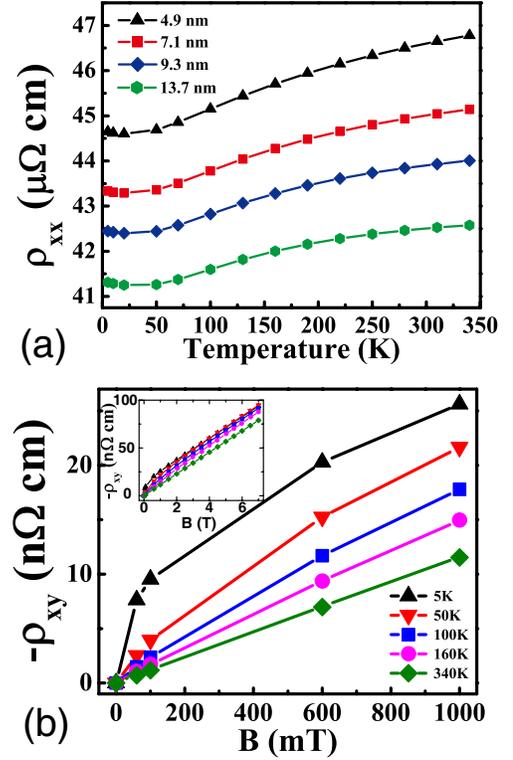}
\caption{(color online) The transport properties of Ni$_{34}$Cu$_{66}$ /MgO(001) film. (a) longitudinal resistivity as a function of temperature for various film thicknesses with no external field, (b) Hall resistivity of the 13.7~nm-thick film as a function of magnetic field for various temperatures. The magnetic field is applied normal to the film plane. The insert shows the entire scope of $\rho_{xy}$ versus $B$ for $B$ up to 7~Tesla. }
\end{figure}

 As a representative case for the series of data at various film thicknesses, the Hall resistivity of $\rho_{xy}$ of 13.7~nm thick Ni$_{0.34}$Cu$_{0.66}$ is shown in Fig. 2(b) as a function of magnetic field at different temperatures. Following the routine procedure to extract the anomalous Hall effect \cite{FeAHE}, we obtain $\rho_{AH0}$ - the corresponding anomalous Hall resistivity at $T=$5 K, together with $\rho_{xx0}$ at $T=$5 K from Fig. 2(a), then plot $\rho_{AH0}/\rho_{xx0}$ versus $\rho_{xx0}$ in Fig. 3(a) as one set of data, together with all the others at various film thicknesses, while its physical implication becomes clear in the following. For such a paramagnetic system at low temperature, because neither the phonon-related contribution nor the intrinsic contribution to the AHE should be considered, one can safely express $\rho_{AH0}$ as the sum of a linear term (the skew scattering) and a quadratic term (the side jump) of $\rho_{xx0}$, i.e.: $\rho_{AH0}=\alpha\rho_{xx0} + \beta\rho_{xx0}^2$, or equivalently
\begin{equation}
\rho_{AH0}/\rho_{xx0}=\alpha + \beta\rho_{xx0}.
\end{equation}
$\alpha$ and and $\beta$ denote the coefficients of the skew scattering and the side jump, respectively. Fig. 3(a) presents the experimentally measured $\rho_{AH0}/\rho_{xx0}$ as a function of $\rho_{xx0}$, which can be well described by a linear function given by the above equation, as shown by the red line in the figure. The intercept and the slope correspond to $\alpha$ and $\beta$ respectively, with the value $\alpha$=$(5.4~\pm~0.2)~\times~10^{-3}$, and $\beta~=~-140~\pm~5$~$\Omega^{-1}$cm$^{-1}$. The finite but nonzero $\beta$ or the quadratic resistivity term is definitely evidenced, for the skew scattering alone apparently cannot explain the experimental data [see the dashed line in Fig. 3(a) of the best fit derived by the least-square method assuming $\beta=$0]. As the intrinsic AHE is by definition ruled out due to the absence of ferromagnetism, the observed quadratic resistivity term directly corresponds to the side jump. With the determined $\alpha$ and $\beta$, the skew scattering contribution and the side jump contribution are shown in Fig. 3(b) by the green and red curves, respectively. Quantitatively, the ratio $|\rho_{AH}^{sj}|/|\rho_{AH}^{sk}|$ is apparently a function of $\rho_{xx0}$ varying from 1.06 as in the thickest film to 1.15 as in the thinnest one. It is noted that the signs of these two contributions remain opposite to each other, meanwhile the amplitude of $\rho_{AH}^{sj}$ is always larger than that of $\rho_{AH}^{sk}$, which leads to a net negative $\rho_{AH}$.

After the direct and unambiguous experimental identification of the pure side jump in the AHE, it is very interesting to further investigate its microscopic origin on this unique platform. Recent experimental study \cite{FeAHE} confirmed that the phonon-induced scattering does not contribute into the skew scattering as suggested theoretically \cite{Bruno2001}. In other word, $\rho_{AH}^{sk}$ is proportional to $\rho_{xx0}$ rather than $\rho_{xx}$. It is yet unclear whether the same conclusion should apply to the side jump mechanism. In fact there are two scenarios existing in the literature concerning this term, i.e. (I) $\rho_{AH}^{sj}~\propto~\rho_{xx0}^2$ as suggested by Ref. \cite{FeAHE,unifiedTheory}, or (II) $\rho_{AH}^{sj}~\propto~\rho_{xx}^2$ as adopted by Ref. \cite{sidejumpexcal,smzhou}. The validity of the two scenarios shall in principle be distinguished by tuning $\rho_{xx}$ via temperature, for the former should be temperature independent but the latter definitely temperature dependent.

\begin{figure}
\centering
\includegraphics[width=8cm]{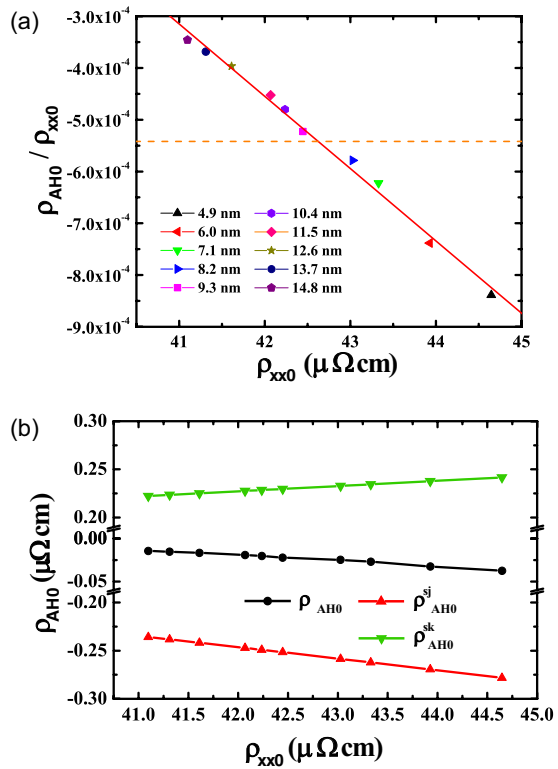}
\caption{ (color online) (a) The plot of $\rho_{AH0}/\rho_{xx0}$  vs $\rho_{xx0}$ obtained at 5~K. The red solid line is the linear fitting employing Eq.~(1). The dashed line is the fitting assuming $\beta=$0. (b) The contribution of side jump $\rho_{AH}^{sj}$ (green) and skew scattering $\rho_{AH}^{sk}$ (red) and the total AHE $\rho_{AH}$ (black) for various film thicknesses at 5~K.}
\end{figure}

Unlike those ferromagnetic systems such as Fe and Ni thin films, in which $M$ almost remains constant in the relevant temperature region where the study of the AHE was conducted \cite{FeAHE,NiAHE}, in paramagnetic systems $M$ certainly varies dramatically with the temperature. The fact that the magnetization $M$ itself is a function of temperature will affect the AHE according to the relation $\rho_{AH}~\propto~M$ \cite{AHEpropM,NagaosaRev2010}.
 For a weak magnetic field, $M(T)$ follows Curie's law, i.e. $M~\propto~1/T$ when $\mu B<<k_B T$. In the following analysis we fix $B$ at 60~mT. $\rho_{AH}$ can be derived by subtracting the normal Hall contribution from the total Hall resistivity, where the normal Hall coefficient is derived by fitting the high field (7~T-5~T) $\rho_{xy}$-$H$ data. Fig. 4(a) shows the temperature dependence of $\rho_{AH}$ for the representative 13.7~nm-thick film when $B=$60~mT, in which a decaying feature as the temperature ascends is clearly seen. In Fig.~4(b) we plot the corresponding $\rho_{AH}$ versus $1/T$ curve. Although for most temperatures explored here $\rho_{AH}$ appears to follow linear dependence on $1/T$ as shown in the insert of Fig.~4(b), it is noted that at the lowest temperatures ($T<20$~K), $\rho_{AH}$ does not strictly follow $M(T)$ or the Curie's law, similar to the previous observation by Fert \emph{et al.} \cite{FertHallrev}. Nevertheless $\rho_{AH}\propto M$ holds for 20~K$<T<$340~K.

\begin{figure}
\centering
\includegraphics[width=8.5cm]{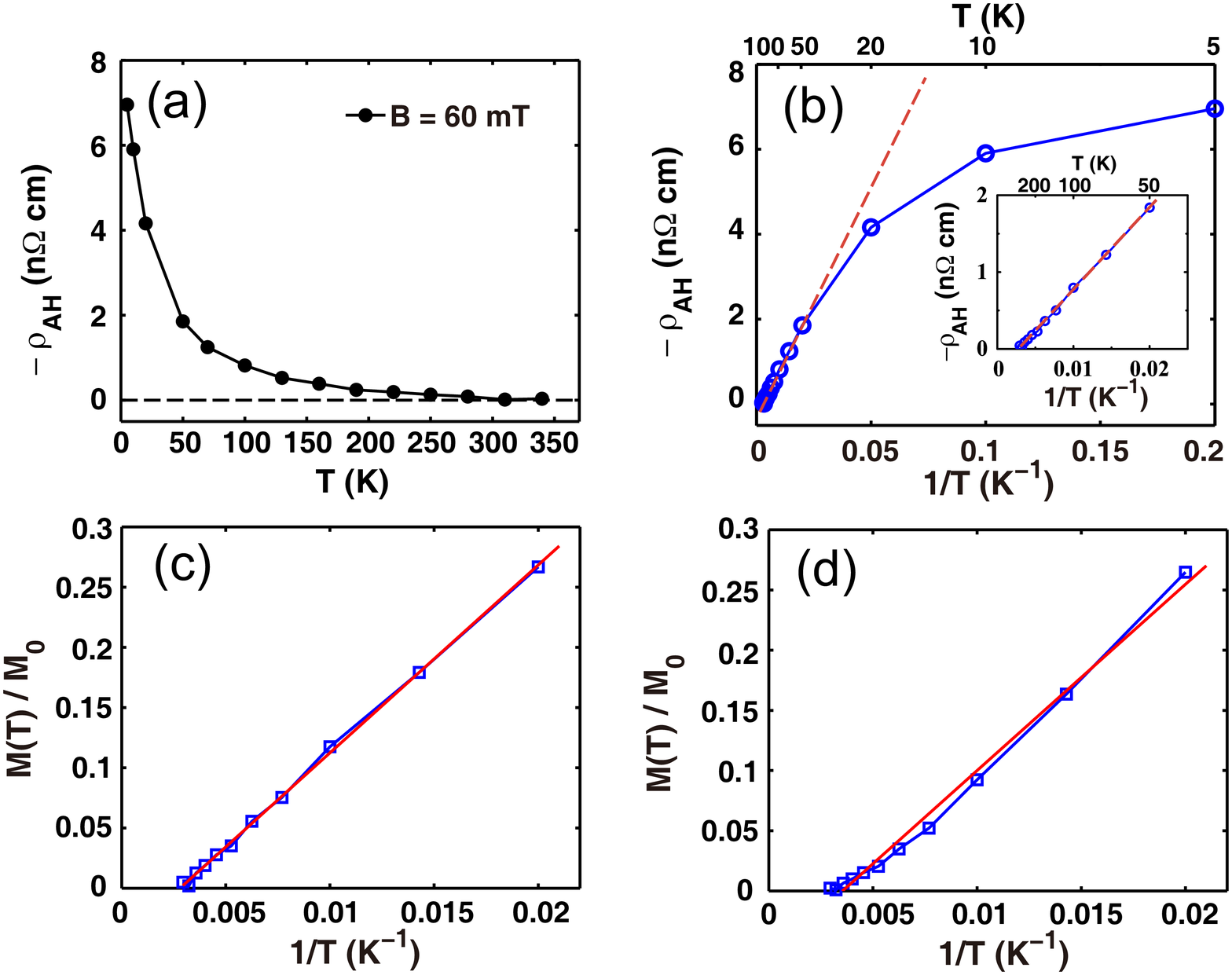}
\caption{ (color online) (a) The temperature dependence of $\rho_{AH}$ and (b)$\rho_{AH}$ vs $1/T$ curve for the 13.7~nm-thick film at $B$=60~mT, derived by subtracting the normal Hall contribution from $\rho_{xy}$. The insert in (b) shows the magnified view for $T>50$~K. The dashed red line is the guide to the eyes.
(c) and (d): $M(T)/M_{5K}$ vs $1/T$ derived from (b) Eq. (\ref{equ:rhoxx0}) and (c) Eq. (\ref{equ:rhoxx}) respectively. The red curves are the fitting results of $M(T)/M_0 \propto k/T$.}
\end{figure}

Considering both the possible scenarios concerning the longitudinal resistivity, Eq. (1) could be extended to a more general expression to reflect the temperature dependence of $M(T)$ explicitly as
\begin{equation}\label{equ:rhoxx0}
\frac{M(T)}{M_0}=\rho_{AH}/(\alpha^\prime M_{0} \rho_{xx0}+\beta^\prime M_{0}\rho_{xx0}^2),    \text{       in case (I)};
\end{equation}
or
\begin{equation}\label{equ:rhoxx}
\frac{M(T)}{M_0}=\rho_{AH}/(\alpha^\prime M_{0} \rho_{xx0}+\beta^\prime M_{0}\rho_{xx}^2),     \text{       in case (II)}.
\end{equation}

Again, the footnote "0" denotes the lowest temperature 5~K as in our experiment. $\alpha^\prime M_{0}$ and $\beta^\prime M_{0}$ are in fact the previously $\alpha$ and $\beta$ respectively that have be derived at low temperature 5~K as presented in Fig. 3(a). Apparently the normalized magnetization $M(T)/M_0$ can be calculated by employing either Eq. (\ref{equ:rhoxx0}) or Eq. (\ref{equ:rhoxx}). The validity of the adapted scenario could be checked by examining whether the correspondingly derived $M(T)$ follows Curie's law, i.e. if $M(T)/M_0$ is linearly proportional to $1/T$. In Fig.~4(c) and Fig.~4(d) we show the normalized magnetization $M(T)/M_0$ of the 13.7~nm-thick film which are derived according to Eq. (\ref{equ:rhoxx0}) and Eq. (\ref{equ:rhoxx}) respectively. While $M(T)/M_{0}$ obtained by assuming $\rho_{AH}^{sj}\propto\rho_{xx0}^2$ shows nice agreement to the linear fitting [Fig.~4(c)]; the assumption $\rho_{AH}^{sj}\propto\rho_{xx}^2$ produces a $M(T)/M_0$ that shows remarkable deviation from the linear fitting [Fig.~4(d)].
This strongly suggests the superiority of Eq. (\ref{equ:rhoxx0}) over Eq. (\ref{equ:rhoxx}). The implication of this sharp contrast is, as we discussed above, that the phonon-induced scattering process does not contribute the side jump mechanism. Although unexpected in pure paramagnetic materials, a negative intercept was found in the linear fitting as presented in Fig. 4(c) and 4(d), similar to the previous observations of a temperature-invariant contribution in the magnetic susceptibility measurement \cite{NiCususp,NiCususp_dilute}. Nevertheless this temperature independent term should not affect our argument.

In conclusion, we have observed a component of the AHE that is proportional to $\rho_{xx0}^2$ in paramagnetic Ni$_{34}$Cu$_{66}$ thin films. With the absence of spin-splitting and hence the intrinsic AHE, the observation of this quadratic term unambiguously points to the presence of the side jump mechanism. Further analysis of the temperature dependence of the AHE points out that the phonon contribution to the side jump is negligible. This implies that elastic scattering processes dominatingly contribute the side jump mechanism.

This work was supported by MOST (No. 2011CB921802) and NSFC (No. 11374057). The technical support from Beijing Synchrotron Radiation Facility (BSRF) for XRD measurements is also acknowledged.

\cleardoublepage

\end{document}